\documentclass{jpconf}
\usepackage{graphicx,bm,amssymb,amsmath,color}
\begin{document}

\title{Real-space perturbation theory for frustrated magnets: application to
magnetization plateaus
}
\author{M E Zhitomirsky}
\address{Service de Physique Statistique, Magn\'etisme et Supraconductivit\'e,
INAC, CEA,\\ 17 rue des Martyrs, 38054 Grenoble Cedex 9, France
}
\ead{mike.zhitomirsky@cea.fr}
\date{\today}

\begin{abstract}
We present a unified approach to the problem of degeneracy lifting in geometrically
frustrated magnets with and without an external field. The method treats fluctuations
around a classical spin configuration in terms of a real-space perturbation expansion.
We calculate two lowest-order contributions for the Heisenberg spin Hamiltonian
and use them to study the magnetization
processes of spin-$S$ triangular and kagom\'e antiferromagnets.
\end{abstract}

\section{Introduction}

A hallmark of geometrical magnetic frustration is degeneracy between lowest-energy
spin configurations that is not related to any underlying symmetry.
Generally, such degeneracy leads to enhanced fluctuations, thermal and quantum.
Fluctuations play different roles depending on their strength. For spin models with
large degeneracy and strong fluctuations, like spin-1/2 kagom\'e and pyrochlore
antiferromagnets, magnetic ordering may be completely suppressed and a spin liquid
state emerges at $T=0$ \cite{Balents10}. On the other hand, if fluctuations are only moderate,
they themselves provide an ordering mechanism commonly known an ``order by disorder effect''
\cite{Villain80,Shender82}.

The analytic tool employed most of the time in theoretical studies of the order by disorder mechanism
consists in the spin-wave expansion for a few reference states, see e.\,g.\ [3\,--\,12].
There is also an alternative route to finding the ground-state selection
that can be loosely called the real-space perturbation theory (RSPT) [13\,--\,18].
Two main advantages of the latter approach are its simple analytic structure, at least in
a few lowest orders, and  description in terms of effective Hamiltonians
operating in the manifold of classical ground states.
This last feature allows to study the effect of fluctuations
without any {\it ad-hoc} assumption about an outcome of the order from disorder
selection. The same method can be also applied for investigation of the order by disorder effect produced
by impurities and bond
disorder \cite{Maryasin13,Maryasin14}.
In Sec.~2 we present a general outline of the real-space perturbation approach
for clean frustrated magnets, which
has been so far missing in the literature. Then, we
apply this theory in Sec.~3 to a prominent problem in frustrated magnetism---width
of the 1/3 magnetization plateau in triangular and kagom\'e antiferromagnets
in an external magnetic field.
The rest of this section is devoted to a  brief discussion of the classical
ground-state constraints in geometrically frustrated magnets with and without an external
magnetic field.

We consider the nearest-neighbor Heisenberg antiferromagnetic model in a 
magnetic field
\begin{equation}
\hat{\cal H} = J \sum_{\langle ij\rangle} {\bf S}_i\cdot{\bf S}_j -
{\bf H}\cdot\sum_i {\bf S}_i \,.
\label{H0}
\end{equation}	
The geometry of frustrated lattices allows decomposition of the nearest-neighbor
spin Hamiltonian (\ref{H0}) into a sum over elementary frustrated units or blocks---triangles,
tetrahedra etc. The minimum-energy condition corresponds, then, to a constraint  on
the total spin ${\bf L}_p= \sum_{i\in p} {\bf S}_i$ of every block. The constraint can be satisfied
in multiple ways, which leads to degeneracy of classical spin configurations at $T=0$.

Let us illustrate block decomposition of the spin Hamiltonian on the example of
the triangular-lattice model. In this case, every exchange bond is shared between two
triangular plaquettes, whereas every spin belongs to six triangles. By including
proper compensation prefactors one can write a sum over spin triangles as
\begin{equation}
\hat{\cal H} = \sum_p  \Bigl[ \frac{J}{4} \,{\bf L}_p^2-\frac{\bf H}{6}\cdot {\bf L}_p
- \frac{J}{4}\bigl( {\bf S}_{1,p}^2 + {\bf S}_{2,p}^2 + {\bf S}_{3,p}^2 \bigr)\Bigr]\,,\quad
{\bf L}_p = {\bf S}_{1,p} + {\bf S}_{2,p} + {\bf S}_{3,p} \,.
\label{H1}
\end{equation}
Minimization with respect to ${\bf L}_p$ yields the ground-state constraint
${\bf L}_p = {\bf H}/(3J)$. The constraint fixes 3 out of 6 angles describing
orientation of three sublattices. An additional continuous parameter is related to the breaking
of the $SO(2)$ rotational symmetry in magnetic field. This leaves in total
2 free parameters, which describe degeneracy of the Heisenberg triangular
antiferromagnet in an external field $0<H<H_s  = 9JS$ \cite{Lee84,Kawamura85}.
For the real-space perturbation expansion we shall need a value of the local magnetic field acting on
an individual spin in a ground-state spin configuration. It is obtained by relaxing for a moment
the fixed spin length and differentiating the ground-state energy:
\begin{equation}
{\bf h}_i = -\frac{\partial E_{\rm g.s.}}{\partial {\bf S}_i} =
\frac{1}{2}\,JS\, {\bf n}_i \sum_p^{i\in p}1  = 3JS {\bf n}_i \,,
\qquad
{\bf n}_i = {\bf S}_i/|{\bf S}_i|\,.
\end{equation}
Note, that only the last term in Eq.~(\ref{H1}) contributes to ${\bf h}_i$
because of the minimum condition with respect to ${\bf L}_p$.
Thus, local fields in a classical ground-state configuration are always parallel
to respective spins and
their strength $H_{loc}=|{\bf h}_i|=3JS$ does not depend on the site index $i$
remaining constant for all fields
$0\leq H\leq H_s$.
It is the site-independence of $H_{loc}$  which
allows to treat all classical ground states on equal footing within the RSPT.

The above approach with only minor modifications applies to many other geometrically
frustrated models \cite{MZ00}.
In particular, for Heisenberg antiferromagnets on kagom\'e and pyrochlore lattices, which
consist of corner-sharing triangles and tetrahedra, the total spin of every block in the
classical ground state is ${\bf L}_p={\bf H}/2J$,
whereas the local field amplitude is $H_{loc}=2JS$. For the nearest-neighbor
Heisenberg antiferromagnet on a face-centered cubic lattice, the spin Hamiltonian is represented
as a sum over edge-sharing tetrahedra with the ground-state constraint ${\bf L}_p={\bf H}/4J$ and
the local field  $H_{loc}=4JS$.

\section{Perturbation expansion around a classical ground state}

\begin{figure}[t]
\hspace*{1pc}
\includegraphics[width=18pc]{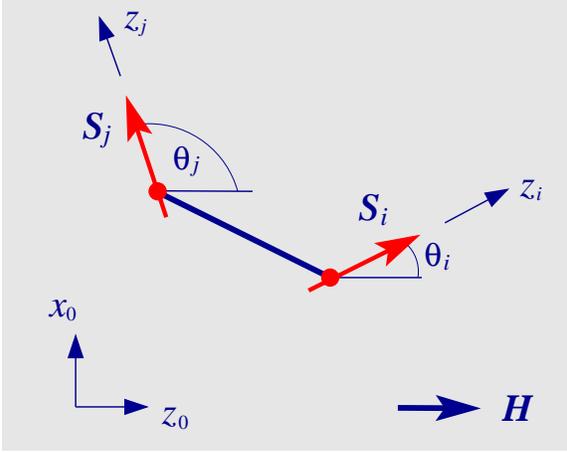}\hspace{2pc}%
\begin{minipage}[b]{14pc}\caption{\label{fig:local}
Choice of local axes and definition of angles for a noncollinear
spin state in an external magnetic field.
}\end{minipage}
\end{figure}

Computation of the classical ground-state energy is equivalent to
the mean-field approximation applied to a quantum spin Hamiltonian.
Corrections to the mean-field approximation can be calculated by treating
perturbatively correlations between spin fluctuations on adjacent sites.
Construction of the perturbation expansion
starts with rewriting the Hamiltonian in the  local spin frame, see Fig.~\ref{fig:local},
and collecting terms that depend on components of only one spin:
\begin{eqnarray}
\hat{\cal H} & = & E_{\rm class} + H_{loc} \sum_i \bigl(S-S_i^z \bigr) +
J \sum_{\langle ij\rangle}  \Bigl[ S_i^yS_j^y + S_i^xS_j^x\cos\theta_{ij}
+ \left(S-S_j^z\right) \left(S-S_j^z\right)\cos\theta_{ij} \nonumber \\
& & \mbox{} + \sin\theta_{ij} \bigl(S_i^xS_j^z-S_i^zS_j^x\bigr)\Bigr]+
H\sum_{i} S_i^x \sin\theta_i \,.
\label{local1}
\end{eqnarray}
The $z_i$-axis on a given site is always pointing along ${\bf S}_i$,
whereas orientation of $x_i$ ($y_i$) is bond-dependent and assumed to
lie in (be orthogonal to) the $z_i$--$z_j$ plane with
$\theta_{ij}$ being an angle between two spins on a given bond.
All terms linear in $S^x_i$ or $S^y_i$ disappear
due to the minimum energy condition. Then, dropping
the classical energy constant $E_{\rm class}$, we obtain
\begin{eqnarray}
&& \hat{\cal H}  =  H_{loc} \sum_i (S-S_i^z)  + \hat{V}_1 + \hat{V}_2 + \hat{V}_3 + \hat{V}_4\,,
\label{V} \\
&& \hat{V}_1 =  -\frac{J}{4} \sum_{\langle ij\rangle} (1-\cos\theta_{ij})
\bigl(S_i^+S_j^+ + S_i^-S_j^- \bigr), \ \
\hat{V}_2  =  \frac{J}{4} \sum_{\langle ij\rangle} (1+\cos\theta_{ij})
\bigl(S_i^+S_j^- + S_i^-S_j^+ \bigr),
\nonumber \\
& &
\hat{V}_3 =
\frac{J}{2}\sum_{i,j}\sin\theta_{ij}(S_j^++S_j^-)(S-S_i^z),
 \qquad\ \
\hat{V}_4 =
J \sum_{\langle ij\rangle} (S-S_i^z)(S-S_i^z) \cos\theta_{ij}\,.
\nonumber
\end{eqnarray}
The first term corresponding to the Zeeman energy in a local field $ H_{loc}$
is chosen as the unperturbed Hamiltonian $\hat{\cal H}_0$ with trivially calculated excited
states, whereas bond terms $\hat{V}_k$  are treated as perturbations.
In principle, there is no an explicit small parameter for doing that.
Still, since $h=O(zJ)$ and $\hat{V}=O(J)$, one can argue that
such an approximation amounts to the $1/z$ expansion, with $z$ being the
coordination number.

In the following we focus on a quantum correction to the ground-state energy.
For analogous RSPT treatment of thermal effects for classical models see
Refs.~\cite{Canals04,Maryasin14}. A noninteracting quantum ground state $|0\rangle$
coincides with a selected classical state and corresponds to a ``fully saturated
state'' in the rotated basis:  $S^+_i |0\rangle= 0$.
The perturbation $\hat{V}$ can be treated using various forms of the
perturbation expansion including the standard Rayleigh-Schr\"odinger theory.
In this way we formulate a few simple rules that are used to identify nonzero contributions:\\
(i) Each term in the perturbation series is represented by a linked cluster ensuring
the correct size scaling $\Delta E \propto N$. Every link corresponds to one of
the perturbation terms  $\hat{V}_k$ acting on a specific lattice bond.
The total number of links is equal to the order of RSPT expansion.
Several links passing the same lattice bond are permitted. \\
(ii) The noninteracting ground state $|0\rangle$ is a vacuum for spin flips.
Therefore, every term in the perturbation series starts and ends with creation
$S^-_iS^-_k$ and annihilation $S^+_{i'}S^+_{k'}$
of a pair of spin flips, corresponding to the action of the $\hat{V}_1$ operator.\\
(iii) All perturbations except of $\hat{V}_3$ conserve parity of the total  number of spin-flips.
Hence, every term in the ground-state energy expansion contains an even number of the $\hat{V}_3$ operators.

\begin{figure}[t]
\center{\includegraphics[width=0.7\columnwidth]{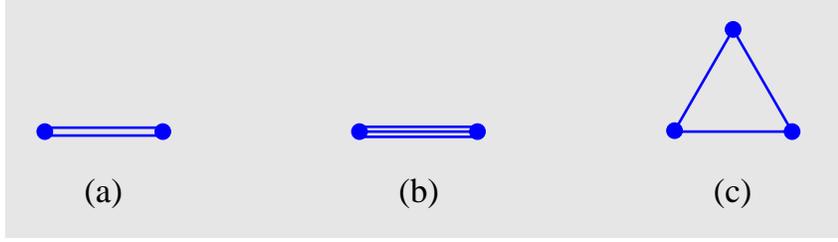}}
\vskip 2mm
\caption{\label{fig:clust}Linked clusters contributing in the second order (a) and in the third order,
(b) and (c), of the real-space perturbation expansion.
}
\end{figure}

We now use the above rules for derivation of second- and third-order RSPT corrections.
According to the  rule (ii), the second-order correction has a very simple form and consists in the double action
of the pair spin-flip operator $\hat{V}_1$ on the same bond:
\begin{equation}
|00\rangle\xrightarrow{S^-_i S^-_j}|11\rangle\xrightarrow{S^+_i S^+_j} |00\rangle \,.
\end{equation}
A graphical representation of this process is shown in Fig.~\ref{fig:clust}(a).
An intermediate state with two noninteracting spin flips has energy $2H_{loc}$.
Calculating matrix elements of spin operators we find
\begin{equation}
E_2  = -\frac{J^2S^2}{8H_{loc}} \sum_{\langle ij\rangle}\, (1-\cos\theta_{ij})^2\,.
\label{dE2}
\end{equation}
Apart from two unimportant terms that
sum up to a state-independent constant,
$E_2$ contains a biquadratic coupling $\sim \cos^2\theta_{ij}=({\bf n}_i\cdot{\bf n}_j)^2$ between nearest-neighbor
spins. The energy (\ref{dE2}) has a meaning of an effective Hamiltonian operating in the manifold
of classical ground states parameterized by $N$ unit vectors ${\bf n}_i$ subject to the constraint.
Therefore, a biquadratic term may arise even for $S=1/2$ frustrated models. The negative sign in front of
the biquadratic coupling favors the ``most collinear'' spin configurations among degenerate classical
ground states. In many models with non-extensive ground state degeneracy,
the order by disorder mechanism selects collinear or coplanar states [3\,--\,8].
This choice can be easily understood on the basis of the effective interaction (\ref{dE2}), i.e.\
without doing any numerical computations that are required in the spin-wave theory.

An effective biquadratic interaction in the most general form (\ref{dE2}) was first published 
by Heinil\"a and Oja \cite{Heinila93} and, by now, has become a part of the verbal tradition in
frustrated magnetism. A natural question to ask in this connection is whether it is
appropriate to describe quantum effects in a frustrated magnet with the help of a biquadratic
term perhaps with a  phenomenological or fitted coefficient [24\,--\,26]. 
Feasibility of such a fit was questioned in Ref.~\cite{Hassan06}, which
showed that the spin-wave energy of the Heisenberg kagom\'e antiferromagnet in a magnetic field
does not follow a simple cosine angular dependence
expected from Eq.~(\ref{dE2}). A similar calculation for an anisotropic $XY$ pyrochlore
antiferromagnet has recently demonstrated that accuracy of
the lowest order RSPT correction may improve significantly with increased anisotropy  \cite{Maryasin14}.
We shall now derive the complete third-order quantum
correction, which has not so far been obtained in the literature and may help to further clarify the
accuracy of Eq.~(\ref{dE2}).

For the frustrated lattices mentioned in Sec.~1, there are two types of third-order
processes corresponding to dimer and triangle clusters that are shown in
Fig.~\ref{fig:clust}. The dimer processes, Fig.~\ref{fig:clust}(b), are represented
by the following diagram:
\begin{equation}
|00\rangle\xrightarrow{S^-_iS^-_j}|11\rangle\xrightarrow{S^z_i S^z_j} |11\rangle
\xrightarrow{S^+_i S^+_j} |00\rangle \,,
\end{equation}
which describes subsequent action of $\hat{V}_1$, $\hat{V}_4$ and again $\hat{V}_1$.
The energy correction from this process is
\begin{equation}
E'_3 =  \frac{J^3S^2}{16H_{loc}^2}\,\sum_{\langle ij\rangle}\, (1-\cos\theta_{ij})^2 \cos\theta_{ij} \,.
\label{dE3a}
\end{equation}
The contribution $E'_3$ describes renormalization of the biquadratic exchange
(\ref{dE2}) by interaction between excited states.

For a  triangular cluster $(ijk)$, Fig.~\ref{fig:clust}(c), a third-order process
is described by the diagram
\begin{equation}
|000\rangle\xrightarrow{S^-_i S^-_j}|110\rangle\xrightarrow{S^+_jS^-_k} |101\rangle
\xrightarrow{S^+_k S^+_i} |000\rangle
\end{equation}
The corresponding energy shift is
\begin{equation}
E_{\triangle} =  \frac{J^3S^3}{32 H_{loc}^2}(1-\cos\theta_{ij})
(1+\cos\theta_{jk})(1-\cos\theta_{ik})e^{i ( \varphi^i_{jk} + \varphi^j_{ik} + \varphi^k_{ji})} \ ,
\end{equation}
where $\varphi^i_{jk}$ is an angle between two planes formed by pairs of spins
$(i,j)$ and $(i,k)$. Technically, the phase factors appear because of bond-dependent
orientation of $x_i$ and $y_i$ axes in Eq.~(\ref{local1}) and, as a result,
in different phase factors of $S_i^{\pm}$  operators in Eq.~(\ref{V}).
Two reverse processes $ij\to jk\to ki$ and $ik\to kj \to ji$
have opposite phases and sum up into a real contribution. The phase factors
disappear for locally coplanar spin structures such that spins on a triangular cluster
form a single plane. Assuming further a coplanar configuration and summing over six possible processes
for a given triangle, we finally obtain
\begin{eqnarray}
E''_3 & = & \frac{J^3S^3}{16H_{loc}^2}\, \sum_{\triangle}\,\Bigl[
(1+\cos\theta_{ij})(1-\cos\theta_{jk})(1-\cos\theta_{ik}) +
(1-\cos\theta_{ij})(1+\cos\theta_{jk})
 \nonumber \\
& & \mbox{\qquad\qquad} \times (1-\cos\theta_{ik})
+ (1-\cos\theta_{ij})(1-\cos\theta_{jk})(1+\cos\theta_{ik})\Bigr].
\label{dE3b}
\end{eqnarray}

Real-space quantum corrections scale with specific powers of $S$. For example,
$E_2, E''_3 \propto JS$ correspond to the harmonic spin-wave approximation, whereas
$E'_3 \propto J$ is a nonlinear contribution. These terms constitute only a part of
the respective spin-wave corrections, which naturally include graphs of arbitrary
length. The number of relevant clusters quickly grows with the order of the real-space
expansion and their evaluation beyond the fourth order may require numerical computations.
Nonetheless,  analytic calculations in higher orders can be used to identify
the lowest-order perturbation processes that lift translational degeneracy for kagom\'e
and pyrochlore antiferromagnets, see, for example, [18,\,27\,--\,29]. 

\section{Fractional magnetization plateaus in frustrated antiferromagnets}

The general expressions for second- and third-order  energy corrections derived in the
previous section can be straightforwardly applied to the problem of order by disorder
selection in magnetic field and, in particular, for calculation of the magnetization
plateau width in triangular and kagom\'e antiferromagnets.

\subsection{Triangular Antiferromagnet}

Degeneracy of the triangular-lattice antiferromagnet in an external field is determined by the classical
constraint ${\bf S}_\triangle = {\bf H}/(3J)$ (Sec.~1), which can be satisfied in multiple
ways by three magnetic sublattices. Among possible spin structures, thermal and quantum fluctuations
select two coplanar states, the Y-state for $H< H_{c1}$ and the $\mathbb{V}$-state for $H> H_{c2}$, and a
collinear $uud$ state for $H_{c1}\leq H\leq H_{c2}$  \cite{Lee84,Kawamura85,Chubukov91},
see Fig.~1. The collinear state is  classically stable
only for a single value of the external field $H_c=H_s/3=3JS$. Quantum fluctuations extend its presence
to a finite range of fields around $H_c$ and produce the $m =1/3$ magnetization plateau. Calculation of the
plateau width within the spin-wave theory is not entirely trivial and  requires some sort
of self-consistent  approximation because of  spurious negative-energy modes arising for the $uud$ state
at $H\neq H_c$, i.e.\ beyond the classical stability point
\cite{Chubukov91,Takano11}.

The problem of nonclassical ground state selection can be readily addressed by studying effective spin
Hamiltonians obtained within the real-space perturbation approach. Using the expressions derived in
Sec.~2, we have checked stability of the collinear spin structure with respect to small canting of
three sublattices. Two critical fields, which bound the plateau region,
are given by
\begin{equation}
H_{c1} = 3JS - \frac{J}{6S}\,, \quad H_{c2} = 3JS + \frac{J}{3}  +  \frac{J}{6S} \,.
\label{Twidth}
\end{equation}
Transitions at the plateau ends are continuous for all values of $S$.
Note that $H_{c1}$ and $H_{c2}$ are shifted asymmetrically with respect to $H_{c}$,
which differs qualitatively from a symmetric relation $H_c=(H_{c1}+H_{c2})/2$
obtained by using only a biquadratic term.

Let us now compare the above analytic expressions to available numerical results on
the width of the 1/3-magnetization plateau in the triangular-lattice  antiferromagnet.
For $S=1/2$, equation (\ref{Twidth}) gives  $H_{c1}=1.167$ and $H_{c2}=2.167$ (in units
of $J$), whereas the exact diagonalization study of finite clusters yields $H_{c1}=1.381$
and $H_{c2}=2.157$ \cite{Honecker04}. For $S=1$, our approximate analytic results are
$H_{c1}=2.833$ and $H_{c2}=3.5$, which should be checked against numerical values
$H_{c1}=2.839$ and $H_{c2}=3.552$ obtained in \cite{Richter13} by combination of the exact
diagonalization and the coupled-cluster methods.  Although such a remarkable
agreement between numerics and a simple analytic theory is partly fortuitous,
the above analysis shows the capability of the real-space perturbation theory
to provide a quantitative description of quantum effects in frustrated magnets.

\begin{figure}[t]
\center{\includegraphics[width=0.85\columnwidth]{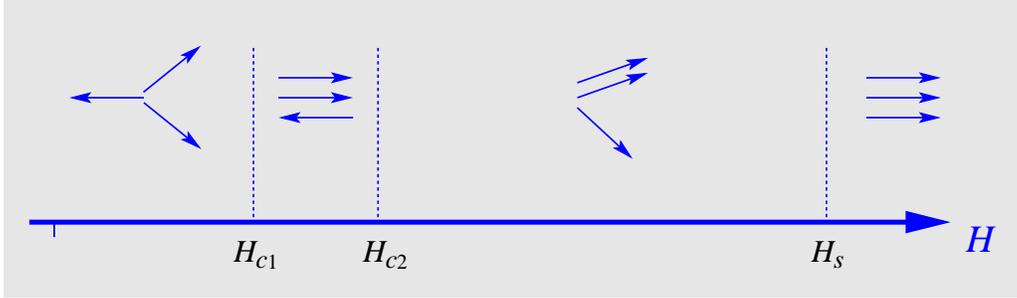}}
\vskip 1mm
\caption{Three-sublattice ground-state spin configurations featured by triangular and kagom\'e
antiferromagnets as a function of an external magnetic field:
the Y-state for $H< H_{c1}$, the $uud$ state for $H_{c1}\leq H\leq H_{c2}$  and
the $\mathbb{V}$-state for $H> H_{c2}$.
}
\label{fig:conf}
\end{figure}

\subsection{Kagom\'e Antiferromagnet}

The kagom\'e antiferromagnet exhibits a much higher degeneracy of the
classical ground states in comparison with the triangular-lattice model.
The difference stems from loose connectivity of triangle blocks in a kagom\'e structure, which form
a network of corner-sharing triangles. As a result, infinitely many states with different
translational patterns have the same classical energy.
Still, at the level of each triangular block, short-range fluctuations select
the same sequence of three-sublattice structures  in a magnetic field
as in the triangular antiferromagnet  \cite{Hassan06,MZ02}.
Here we consider only this `short-range' part of the quantum order by disorder effect
in the kagom\'e antiferromagnet, because it is a much stronger effect than selection
of a specific translation pattern and at the same time is responsible for the $m = 1/3$ magnetization
plateau. Accordingly, we do not discuss nature of the plateau state:  it can either possess some type of
long-range order or remain in a spin liquid state.

Analysis of the combined contribution $E_2+E'_3+E''_3$ shows that the $uud$ state is stable
in a finite window of fields around $H_c=2JS$:
\begin{equation}
2JS - \frac{J}{8} -\frac{J}{4S} \leq H \leq 2JS + \frac{3J}{8}  +  \frac{J}{4S} \,.
\label{Kwidth}
\end{equation}
In the extreme quantum case of $S=1/2$, the width of the magnetization plateau obtained
from Eq.~(\ref{Kwidth}) is too large compared to available numerical results \cite{Capponi13,Nishimoto13}.
On the other hand, for $S=1$, our approximate calculation yields
$H_{c1}=1.625$ and $H_{c2}=2.625$, which stand reasonably well against numerical DMRG values
$H_{c1}\approx 1.67$ and $H_{c2}\approx 2.68$ \cite{Picot14}.
Deficiency of the low-order RSPT calculation for the $S=1/2$ kagom\'e antiferromagnet is not very
surprising due to neglect of the tunneling processes between classical ground states
that appear only in higher orders of the real-space expansion. The RSPT approach also fails to
predict other fractional magnetization plateaus with $m = 5/9$  and 7/9 determined by coherent
hopping of spin flips around hexagons \cite{Capponi13}.
Still, a reasonable match between analytic and numerical values of the critical fields for $S=1$
suggests a capability of the RSPT method to treat quantitatively even strongly frustrated spin models with
$S>1/2$.

\section{Conclusions}

We have given a general outline of the real-space perturbation theory for frustrated
Heisenberg antiferromagnets. With little modifications the above approach also works
for frustrated models with anisotropic exchange, see e.\,g.~\cite{Maryasin14}.
However, for the single-ion anisotropy, quantum fluctuations are already present at the level
of the noninteracting Hamiltonian $\hat{\cal H}_0$ and this case requires a separate
analysis. Let us also mention here that a somewhat different perturbation scheme was used
by the authors of Ref.~\cite{Bergman07}. They have combined the local field term with
the Ising part of the spin-flip interaction in Eq.~(\ref{V}) into a new noninteracting
Hamiltonian $\hat{\cal H}_0+\hat{V}_4$. This is equivalent to partial resummation of
the perturbation series. Though, such resummation may be helpful for a certain class
of problems,  one disadvantage of this procedure is that
the linked cluster representation of different perturbation terms is lost and one has
to distinguish contractible and non-contractible clusters \cite{Bergman07}.

The RSPT expansion provides a simple qualitative
description of the effect of quantum and thermal fluctuations on the ground state
selection in geometrically frustrated magnets. The question of quantitative accuracy
of the low-order RSPT results is, however, more delicate. Reasonable agreement
between the third-order RSPT and numerical results for the plateau width in $S=1$ triangular
and kagom\'e antiferromagnets does not immediately imply that a next order correction would
further improve the agreement. Nonetheless,
the RSPT expansion carried out numerically to high orders by analogy with the standard
Ising and dimer series expansion techniques for nonfrustrated magnets \cite{Gelfand00,Zheng06,Jackeli07}, 
should be able to address quantitatively many open questions and problems in the field of frustrated magnetism.

\ack
I am grateful to 
Benjamin Canals, Sasha Chernyshev, George Jackeli, Vladimir Maryasin and Roderich Moessner
for numerous discussions of questions raised in this work and to Andreas Honecker for careful reading
of the manuscript.


\section*{References}
\begin{thebibliography}{99}

\bibitem{Balents10}
Balents L 2010 \textit{Nature} {\bf 464} 199

\bibitem{Villain80}
Villain J,  Bidaux R, Carton J-P and Conte R 1980
\textit{J. de Physique} \textbf{41} 1263

\bibitem{Shender82}
Shender E F 1982 \textit{Sov. Phys. JETP} \textbf{56} 178

\bibitem{Kawamura84}
Kawamura H 1984 \textit{J. Phys. Soc. Jpn.} \textbf{53} 2452

\bibitem{Henley87}
Henley C L 1987 \textit{J. App. Phys.} \textbf{61} 3962

\bibitem{Oja87}
Vierti\"o H E and Oja A S 1987
\textit{Phys. Rev.} B {\bf 36} 3805

\bibitem{Henley89}
Henley C L 1989 \textit{Phys. Rev. Lett.} \textbf{62} 2056


\bibitem{Chubukov91}
Chubukov A V and Golosov D I 1991 \textit{J. Phys.: Condens. Matter} \textbf{3} 69

\bibitem{Ritchey93}
Ritchey I, Chandra P and Coleman P 1993
\textit{Phys. Rev.} B \textbf{47} 15342(R)

\bibitem{Harris01}
Harris A B,  Aharony A, Entin-Wohlman O, Korenblit I Y, Birgeneau R J and  Kim Y-J 2001
\textit{Phys. Rev.} B \textbf{64} 024436

\bibitem{Hassan06}
Hassan S R and Moessner R 2006
\textit{Phys. Rev.} B \textbf{73} 094443

\bibitem{Zhitomirsky12}
Zhitomirsky M E, Gvozdikova M V, Holdsworth P C W and Moessner R 2012
\textit{Phys. Rev. Lett.} \textbf{109} 077204

\bibitem{Lindgard88}
Lindg\aa rd P-A 1988
\textit{Phys. Rev. Lett.} \textbf{61} 629

\bibitem{Long89}
Long M W 1989 \textit{J. Phys.: Condens. Matter} \textbf{1} 2857

\bibitem{Heinila93}
Heinil\"a M T and Oja A S 1993
\textit{Phys. Rev.} B {\bf 48} 7227

\bibitem{Zhang02}
Zhang N-G, Henley C L, Rischel C and Lefman K 2002
\textit{Phys. Rev.} B {\bf 65} 064427

\bibitem{Canals04}
Canals B and Zhitomirsky M E 2004
\textit{J. Phys.: Condens. Matter} {\bf 16} S759

\bibitem{Bergman07}
Bergman D L, Shindou R, Fiete G A  and Balents L 2007
\textit{Phys. Rev.} B {\bf 75} 094403

\bibitem{Maryasin13}
Maryasin V S and Zhitomirsky M E 2013
{\it Phys. Rev. Lett.} {\bf 111} 247201

\bibitem{Maryasin14}
Maryasin V S and Zhitomirsky M E 2014
{\it Phys. Rev.} B {\bf 90} 094412

\bibitem{Lee84}
Lee D H, Joannopoulos J D, Negele J W  and Landau D P 1984
\textit{Phys. Rev. Lett.} {\bf 52} 433

\bibitem{Kawamura85}
Kawamura H and Miyashita S 1984 \textit{J. Phys. Soc. Jpn.} \textbf{54} 4530

\bibitem{MZ00}
Zhitomirsky M E, Honecker A and Petrenko O A 2000
\textit{Phys. Rev. Lett.} {\bf 85} 3269

\bibitem{Nikuni98}
Nikuni T and Jacobs A E 1998
{\it Phys. Rev.} B {\bf 57} 5205

\bibitem{Larson08}
Henley C L 2001 \textit{Can. J. Phys.} {\bf 79} 1307

\bibitem{Griset11}
Griset C, Head S, Alicea J and Starykh O A 2011
\textit{Phys. Rev.} B \textbf{84} 245108

\bibitem{Cabra05}
Cabra D C, 
Grynberg M D, Holdsworth P C W, Honecker A, Pujol P,
Richter  J, Schmalfuss D and Schulenburg J 2005
\textit{Phys. Rev.} B {\bf 71} 144420

\bibitem{Hizi09}
Hizi U and Henley C L 2009
\textit{Phys. Rev.} B {\bf 80} 014407

\bibitem{Chern14}
Chernyshev A L and Zhitomirsky M E 2014
\textit{Phys. Rev. Lett.} {\bf 113} 237202

\bibitem{Takano11}
Takano J, Tsunetsugu H and Zhitomirsky M E 2011
\textit{J. Phys.: Conf. Series} {\bf 320} 012011

\bibitem{Honecker04}
Honecker A, Schulenburg J and Richter J 2004
\textit{J. Phys.: Condens. Matter} \textbf{16}  S749

\bibitem{Richter13}
Richter J, G\"otze O,  Zinke R,  Farnell D J J and Tanaka H 2013
\textit{J. Phys. Soc. Jpn.} \textbf{82} 015002

\bibitem{MZ02}
Zhitomirsky M E 2002
\textit{Phys. Rev. Lett.} {\bf 88} 057204

\bibitem{Capponi13}
Capponi S, Derzhko O, Honecker A, L\"auchli A M and Richter J 2013
\textit{Phys. Rev.} B {\bf 88}, 144416

\bibitem{Nishimoto13}
Nishimoto S, Shibata N and Hotta C 2013
\textit{Nature Comm.} {\bf 4} 2287

\bibitem{Picot14}
Picot T and Poilblanc D 2014
\textit{Preprint} arXiv:1406.7205

\bibitem{Gelfand00}
Gelfand M P and Singh R R P 2000 \textit{Avd. Phys.} {\bf 49} 93

\bibitem{Zheng06}
Zheng W, Fj\ae restad J O, Singh R R P, McKenzie R H and Coldea R 2006
\textit{Phys. Rev.} B {\bf 74} 224420

\bibitem{Jackeli07}
Jackeli G and Ivanov D A 2007
\textit{Phys. Rev.} B {\bf 76} 132407


\end{thebibliography}
\end{document}